\documentclass[a4paper,11pt]{article}
\usepackage{pos}
\usepackage{framed}
\usepackage{listings}
\usepackage{slashed}
\input{let-color.sty}
\newcommand{\marty}{\emph{MARTY}}
\newcommand{\csl}{\emph{CSL}}
\newcommand{\martyv}[1]{\emph{MARTY-#1}}
\title{Semi-automated BSM model building procedures in \martyv{1.1} through a 2HDM example}
\ShortTitle{Model Building with MARTY}
\author*[a]{G. Uhlrich}
\author[a,b]{F. Mahmoudi}
\author[a,b,c]{A. Arbey}

\affiliation[a]{Universit\'e de Lyon, Universit\'e Claude Bernard Lyon 1, CNRS/IN2P3, \\
Institut de Physique des 2 Infinis de Lyon, UMR 5822, F-69622, Villeurbanne, France}

\affiliation[b]{Theoretical Physics Department, CERN, CH-1211 Geneva 23, Switzerland}
\affiliation[c]{Institut Universitaire de France, 103 boulevard Saint-Michel, 75005 Paris, France}

\emailAdd{g.uhlrich@ipnl.in2p3.fr}
\emailAdd{nazila@cern.ch}
\emailAdd{alexandre.arbey@ens-lyon.fr}

\abstract{\marty\ is a C++ computer algebra system specialized for High Energy Physics that can calculate amplitudes, squared amplitudes and Wilson coefficients in a large variety of beyond the Standard Model scenarios up to the one-loop order. It is fully independent of any other framework and its main development guideline is generality, in order to be adapted easily to any type of model. The calculations are fully automated from the Lagrangian up to the generation of the C++ code evaluating the theoretical results (numerically, depending on the model parameters). Once a phenomenological tool chain has been set up - from a Lagrangian to observable analysis - it can be used in a model independent way leaving only model building, with \marty, as the task to be performed by physicists. Here we present the main steps to build a general new physics model, namely gauge group, particle content, representations, replacements, rotations and symmetry breaking, using the example of a 2 Higgs Doublet Model. The sample codes that are shown for this example can be easily generalized to any Beyond the Standard Model scenario written with \marty.}

\FullConference{%
  Tools for High Energy Physics and Cosmology - TOOLS2020\\
  2-6 November, 2020\\
  Institut de Physique des 2 Infinis (IP2I), Lyon, France}

\begin{document}
\maketitle

\section{Introduction}

\marty\ is an open source code for symbolic computation in high-energy physics~\cite{marty} and can be downloaded from its \href{https://marty.in2p3.fr}{website}\footnote{\url{https://marty.in2p3.fr}} where all related publications, manuals and documentations can be found. Symbolic computations are done using the \csl\ module of \marty, a C++ Symbolic computation Library dedicated to this task. We present some of the \csl\ features in the following but extensive information can be found in the \href{https://marty.in2p3.fr/doc/csl-manual.pdf}{CSL manual}. 

The 2 Higgs Doublet Model (2HDM) is one of the simplest extension of the Standard Model (SM) as it contains only one additional representation, namely another Higgs doublet (four real degrees of freedom) in the $SU(2)_L$ group. In the SM, the gauge symmetry breaking $SU(2)_L\times U(1)_Y\to U(1)_{em}$ of the Higgs doublet $\Phi_1$ gives masses to the weak bosons (three longitudinal polarizations for $W^\pm$ and $Z$) leaving one real degree of freedom, the so-called Higgs boson $h^0$. When adding a second Higgs doublet $\Phi_2$ in the Lagrangian four real degrees of freedom acquire masses: another neutral scalar $H^0$, a charged scalar $H^\pm$ and a pseudo-scalar $A^0$.

\begin{figure}[h!]
    \centering
    \includegraphics[width=0.55\linewidth]{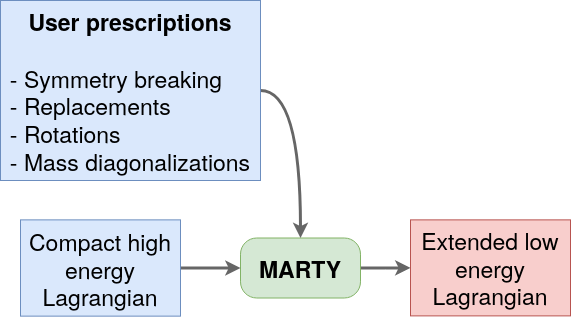}
    \caption{Semi-automated model building concept in \marty.}
    \label{fig:semiauto}
\end{figure}

Two main strategies are possible to build a Beyond the Standard Model (BSM) model in \marty. The first one is to give the Lagrangian term by term, that is straight-forward but becomes long and error prone for more evolved BSM models. The second strategy is to use \marty's top-down machinery for model building: given a high energy Lagrangian with all symmetries preserved - much more compact - and appropriate prescriptions, the low energy Lagrangian can be derived by \marty. It is semi-automated because users have to define the different steps of model building, but each one uses built-in \marty\ functions that are fully automated. This strategy represented in Figure \ref{fig:semiauto} reduces as much as possible the information that has to be given by the user, and thus the number of mistakes that can be introduced in the model.

For the model definition and conventions, we follow Ref.~\cite{2HDM}. The SM flavour is ignored here for simplicity but the built-in 2HDM included in \marty contains the three fermion families. We treat therefore only one generation of fermions here.
 
The order of the different model building steps is standardized in \marty. Section \ref{sec:gauge} shows how to initialize gauge and flavour groups. Particle content is introduced in Section \ref{sec:content} and custom interaction terms are added in Section \ref{sec:lagrangian}. Symmetry breaking and rotation to the mass eigenstate basis are demonstrated respectively in Sections \ref{sec:breaking} and \ref{sec:rotation}, and finally Section \ref{sec:completion} presents how to complete the model building procedure. All the code samples presented in the following are specific to 2HDM, but can be easily generalized to any BSM model. More information on model building is given in the dedicated chapter of the \href{https://marty.in2p3.fr/doc/marty-manual.pdf}{MARTY manual}.

\section{The unbroken Lagrangian}

\subsection{Gauge and Flavour}
\label{sec:gauge}
The 2HDM lies in the same gauge as the unbroken SM: $SU(3)_C\times SU(2)_L\times U(1)_Y$. For gauged groups the type\footnote{There is no $U(N)$ in \marty, only $U(1)$ and $SU(N)$ as $\mathfrak{u}(N)\cong \mathfrak{su}(N)\oplus \mathfrak{u}(1)$. A $U(N)$ group can then be coded by introducing a $SU(N)$ and a $U(1)$.} must be chosen amongst \lstinline!U1, SU, SO, Sp, E6, E7, E8, F4, G2! in the namespace \lstinline!group::Type!. For the sake of simplicity in this write-up we do not add the SM flavour group of dimension 3, we thus have only one fermion generation and no lepton or quark flavour. Gauge group initialization is presented in the sample code below.

\begin{framed}
\begin{lstlisting}[language=C++]
    Model THDM;
    THDM.addGaugedGroup(group::Type::SU, "C", 3); // SU(3)_C, coupling g_C, boson A_C
    THDM.addGaugedGroup(group::Type::SU, "L", 2); // SU(2)_L, coupling g_L, boson A_L
    THDM.addGaugedGroup(group::Type::U1, "Y"); // U(1)_Y, coupling g_Y, boson A_Y
    THDM.init();
\end{lstlisting}
\end{framed}

Once the gauge and flavour groups have been set, the model must be initialized with the \lstinline!init()! function before adding particles. 

\subsection{Particle content}
\label{sec:content}

The particle content of the model is presented in Table \ref{tab:content}.
\begin{table}[h]
    \centering
    $$\begin{array}{c|c|c|c}
         \text{Particle}& SU(3)_C &SU(2)_L&U(1)_Y\\\hline
         \Phi _1  & 1& 2& 1/2\\
         \Phi _2  & 1& 2& 1/2\\
         E_L      & 1& 2& -1/2\\
         e_R      & 1& 1& -1\\
         U_L      & 3& 2& 1/6\\
         u_R      & 3& 1& 2/3\\
         d_R      & 3& 1& -1/3\\
    \end{array}$$
    \caption{Particle content of the 2HDM without flavour. Irreducible representations are given for the three gauged groups by dimension or $U(1)-$charge. There is one fermion generation and the two Higgs doublets $\Phi_1$ and $\Phi_2$.}
    \label{tab:content}
\end{table}

Adding a particle to \marty\ is done through three steps. First creating the particle choosing its spin (scalar, fermion, vector boson etc), as presented below.
\begin{framed}
\begin{lstlisting}
    Particle Phi_1 = scalarboson_s("Phi_1", THDM); 
    Particle U_L = weylfermion_s("U_L", THDM, Chirality::Left);
    Particle u_R = weylfermion_s("u_R", THDM, Chirality::Right);
    Particle d_R = weylfermion_s("d_R", THDM, Chirality::Right);
\end{lstlisting}
\end{framed}
Then, non-trivial representations (gauge or flavour) must be set specifying Dynkin labels \cite{dynkin} for non Abelian gauged groups, and the fractional charge for $U(1)$. 
\begin{framed}
\begin{lstlisting}
    Phi_1->setGroupRep("L", {1}); // 1 is doublet SU(2) representation
    Phi_1->setGroupRep("Y", {1, 2}); // 1/2 charge
    U_L->setGroupRep("C", {1, 0}); // (1, 0) is triplet SU(3) representation
    U_L->setGroupRep("L", {1}); 
    U_L->setGroupRep("Y", {1, 6}); // 1/6 charge
    u_R->setGroupRep("C", {1, 0}); 
    u_R->setGroupRep("Y", {2, 3}); 
    d_R->setGroupRep("C", {1, 0}); 
    d_R->setGroupRep("Y", {-1, 3}); 
\end{lstlisting}
\end{framed}
To know the set of Dynkin labels which corresponds to a particular representation, one can check the group theory chapter of the \href{https://marty.in2p3.fr/doc/marty-manual.pdf}{manual}, where common representations for all semi-simple groups are presented together with a general method, using a built-in feature, to derive easily all the other representations if needed.

Finally adding the particle to the model, standard kinetic and gauge interaction terms are introduced automatically by \marty\ at this point.
\begin{framed}
\begin{lstlisting}
    THDM.addParticles({Phi_1, U_L, u_R, d_R});
\end{lstlisting}
\end{framed}
The other particles presented in Table \ref{tab:content} can be defined straight-forwardly in a similar way.

\subsection{Additional interaction terms}
\label{sec:lagrangian}
In the 2HDM some Lagrangian terms do not derive from gauge interactions and must therefore be user-defined. These terms are Yukawa couplings 
\begin{equation}
\label{eq:yukawa}
\begin{split}
    \mathcal{L}_{Yukawa}=
    &-Y_e^{(1)}\bar{E}_L e_R \Phi_ 1 -Y_e^{(2)}\bar{E}_L e_R \Phi_ 2 + \text{h.c.}\\
    &-Y_u^{(1)}\bar{U}_L u_R \tilde{\Phi}_1 -Y_u^{(2)}\bar{U}_L u_R \tilde{\Phi}_ 2 + \text{h.c.}\\
    &-Y_d^{(1)}\bar{U}_L d_R \Phi_ 1 -Y_d^{(2)}\bar{U}_L d_R \Phi_ 2 + \text{h.c.},
\end{split}
\end{equation}
with 
\begin{equation}
\label{eq:tilde}
    \tilde{\Phi}_i\equiv i\sigma^2\Phi_i^\dagger = \epsilon \Phi_i^\dagger,
\end{equation}
and the Higgs potential (see Eq. (2) in Ref.~\cite{2HDM})
\begin{equation}
\label{eq:higgspot}
\begin{split}
-V_{\Phi _1,\Phi _2} =\,&m_{11}^2\Phi _1^\dagger \Phi _1 +  m_{22}^2\Phi _2^\dagger \Phi _2 -m_{12}^2\left(\Phi _1^\dagger \Phi _2 + \Phi _2^\dagger \Phi _1\right)+\dfrac{\lambda _1}{2}\left(\Phi _1^\dagger\Phi _1\right)^2+\dfrac{\lambda _2}{2}\left(\Phi _2^\dagger\Phi _2\right)^2\\
+&\lambda _3\left(\Phi _1^\dagger\Phi _1\right)\left(\Phi _2^\dagger\Phi _2\right) +\lambda _4\left(\Phi _1^\dagger\Phi _2\right)\left(\Phi _1^\dagger\Phi _2\right)+\dfrac{\lambda _5}{2}\left[\left(\Phi _1^\dagger\Phi _2\right)^2+\left(\Phi _2^\dagger\Phi _1\right)^2\right].
\end{split}
\end{equation}

\subsubsection{Initializing content for symbolic expressions}
Before adding terms to the Lagrangian one has to create the required building blocks, i.e. constants and indices for the \csl\ symbolic machinery. First real Higgs parameters:
\begin{framed}
\begin{lstlisting}
    Expr m_11 = constant_s("m_11"); // Same for m_12, m_22
    Expr l_1 = constant_s("l_1"); // Same for l_2, l_3, l_4, l_5
\end{lstlisting}
\end{framed}
Then the Yukawa couplings:\footnote{We take Yukawas to be real here as they will simply be identified with $\sqrt{2}\frac{m_f}{v}$ for a fermion of mass $m_f$ coupled to a Higgs with a Vacuum Expectation Value $v$.}
\begin{framed}
\begin{lstlisting}
    Expr Y_e = constant_s("Y_e");// Same for Y_u, Y_d 
\end{lstlisting}
\end{framed}
And finally creating Dirac and gauge indices: 
\begin{framed}
\begin{lstlisting}
    Index a = DiracIndex();
    Index A = GaugeIndex(THDM, "C", "U_L"); // SU(3) triplet index
    Index i = GaugeIndex(THDM, "L", "E_L"); // SU(2) doublet index
    Index j = GaugeIndex(THDM, "L", "E_L"); // SU(2) doublet index
\end{lstlisting}
\end{framed}
In order to get gauge indices from a model, a group and a particle name must be given to specify in which representation the index must be generated. For the indices \lstinline!i! and \lstinline!j! any $SU(2)_L$ doublet (\lstinline!E_L, U_L, Phi_1, Phi_2!) can be given as they lie in the same representation, the generated indices can then be used for any of those fields. Finally, we need the $\epsilon$ tensor in the $SU(2)_L$ doublet space to build $\tilde{\Phi}$ that was defined in Eq.~(\ref{eq:tilde}). This is done with the following instruction:
\begin{framed}
    \begin{lstlisting}
    Tensor eps = GetEpsilon(GetVectorSpace(THDM, "L", "E_L"));
    \end{lstlisting}
\end{framed}

\subsubsection{Yukawa couplings}
Without the SM flavour symmetry, Yukawa couplings are not $3\times 3$ matrices, simply scalar constants. Equation (\ref{eq:yukawa}) presents these couplings in full generality but fermions are usually defined to couple to only one of the two doublets. This is presented in Table \ref{tab:2HDMType}, following conventions defined in Ref.~\cite{2HDM}. In the following we consider the 2HDM of type IV, lepton specific.
\begin{table}[!h]
	\centering
	\begin{tabular}{l|c c c c}
		Type                & $u_R$ & $d_R$ & $e_R$\\\hline
		I    				&   +   &   +   &   +  \\
		II  			    &   +   &   -   &   -  \\
		III / Flipped       &   +   &   -   &   +  \\
		IV / Lepton specific&   +   &   +   &   -  
	\end{tabular}
	\caption{\label{tab:2HDMType}The different types of 2HDM are defined by the $Z_2$ charges of right-handed fermions specifying to which scalar they couple to in Yukawa terms: $\Phi _1$ (charge -1) or $\Phi _2$ (charge +1).}
\end{table}

The Yukawa Lagrangian can be built in \marty\ using the sample code below.
\begin{framed}
    \begin{lstlisting}
        THDM.addLagrangianTerm(
            -Y_e * cc(E_L({i, a})) * e_R(a) * Phi_1(i),
            true); // Add also the Hermitian conjugate
        THDM.addLagrangianTerm(
            -Y_u * cc(U_L({A, i, a})) * u_R({A, a}) * eps({i, j}) * cc(Phi_2(j)),
            true); // Add also the Hermitian conjugate
        THDM.addLagrangianTerm(
            -Y_d * cc(U_L({A, i, a})) * d_R({A, a}) * Phi_2(i),
            true); // Add also the Hermitian conjugate
    \end{lstlisting}
\end{framed}
\noindent We abbreviated the \lstinline!GetComplexConjugate()! \csl\ function into \lstinline!cc()!. For a \marty\ fermion, the $\gamma^0$ matrix is implicit and 
\begin{equation}
    \psi^* \equiv \bar{\psi} = \psi^*\gamma^0.
\end{equation}
Indices are always ordered in the same way in quantum tensor fields:
\begin{itemize}
    \item Flavour indices in the order of group definition.
    \item Gauge indices in the order of group definition.
    \item Space-time indices, Dirac or Minkowski for fermions or vector bosons respectively.
\end{itemize}
If only one index has to be given, the curly braces \lstinline!{}! can be omitted.

\subsubsection{The Higgs potential}
The Higgs potential can now be defined easily, first defining the four scalar quantities appearing in the potential\footnote{Again using the abbreviation \lstinline!GetComplexConjugate() -> cc()!.}:
\begin{framed}
\begin{lstlisting}
    Expr s_11 = cc(Phi_1(i)) * Phi_1(i);
    Expr s_12 = cc(Phi_1(i)) * Phi_2(i);
    Expr s_21 = cc(Phi_2(j)) * Phi_1(j);
    Expr s_22 = cc(Phi_2(j)) * Phi_2(j);
\end{lstlisting}
\end{framed}
Then adding one by one the terms of the Higgs potential:
\begin{framed}
\begin{lstlisting}
    THDM.addLagrangianTerm(-m_11*m_11 * s_11); // same for phi2
    THDM.addLagrangianTerm(-m_12*m_12 * (s_12 + s_21));
    THDM.addLagrangianTerm(-l_1/2 * pow_s(s_11, 2)); // same for phi2
    THDM.addLagrangianTerm(-l_3 * s_11*s_22);
    THDM.addLagrangianTerm(-l_4 * s_12*s_21);
    THDM.addLagrangianTerm(-l_5/2 * (pow_s(s_12, 2) + pow_s(s_21, 2)));
\end{lstlisting}
\end{framed}
The indices are treated carefully in order to avoid conflicts, as more than two identical indices in a \csl\ expression is ill-defined. The \lstinline!pow_s()! \csl\ function takes care of renaming the indices when expanding, and for regular products we ensure to have two different pairs of indices \lstinline!i! and \lstinline!j! in the definitions of \lstinline!s_11!, \lstinline!s_12!, \lstinline!s_21! and \lstinline!s_22!.

\section{Symmetry breaking}
\label{sec:breaking}

\subsection{Gauge symmetry breaking}
Once the high energy Lagrangian is implemented, we can break it to make \marty\ derive the low energy one. First, one can break the $SU(2)_L\times U(1)_Y$ gauge symmetry by typing: 
\begin{framed}
\begin{lstlisting}
    THDM.breakGaugeSymmetry("Y");
    THDM.breakGaugeSymmetry(
        "L", // Broken group
        {"Phi_1", "Phi_2", "A_L", "U_L", "E_L"}, // Unbroken names
        {{"Phi_11", "Phi_12"}, // Broken names for Phi_1
         {"Phi_21", "Phi_22"}, // Broken names for Phi_2
         {"W_1", "W_2", "W_3"}, // Broken names for A_L
         {"u_L", "d_L"}, // Broken names for U_L
         {"\\nu_L ", "e_L"}}); // Broken names for E_L
\end{lstlisting}
\end{framed}
Names can be omitted, but we prefer here to specify them in order to recover standard conventions.

\subsection{Expansion around Higgs' VEVs}

Minimizing the Higgs potential of Eq.~(\ref{eq:higgspot}) can be done by searching for the classical solutions for $\Phi_i$: 
\begin{equation}
\left\lbrace\begin{array}{l}
\dfrac{\partial V_{\Phi _1, \Phi _2}}{\partial \Phi _i^\dagger} = 0,\\
\dfrac{\partial ^2V_{\Phi _1, \Phi _2}}{\partial \Phi _i\partial \Phi _i^\dagger} > 0.
\end{array}\right.
\end{equation}
Solutions can be expressed without loss of generality as
\begin{equation}
\Phi _i = \left(\begin{array}{c}
0 \\ \dfrac{v_i}{\sqrt{2}}\end{array}\right).
\end{equation}

In terms of quantum mechanics, we interpret this minimum as the Vacuum Expectation Value (VEV) of the Higgs doublets $\langle\Phi _1\rangle$ and $\langle\Phi _2\rangle$. We then expand around this minimum to get degrees of freedom (dof) centered on zero. In the 2HDM it is convenient to define
\begin{equation}
\label{eq:Phi_def}
\Phi _i = \left(\begin{array}{c}
\Phi _{i1}\\
\Phi _{i2}
\end{array}\right) \equiv \left(\begin{array}{c}
\phi _i^+\\
\left(v_i + \rho _i + i\eta _i\right) / \sqrt{2}
\end{array}\right).
\end{equation}
When breaking the gauge symmetry with \marty, $\Phi_i$ fields are broken in two pieces each, i.e. the two components of their doublets. These particles are created automatically by \marty\ and must then be obtained from the model (given their names), and new particles $\phi _i^+, \rho _i$ and $\eta_i$ must be created to replace them. The following sample code demonstrates this procedure for the first doublet, first getting all the required \lstinline!Particle! objects:
\begin{framed}
\begin{lstlisting}
    Particle Phi_11 = THDM.getParticle("Phi_11");
    Particle Phi_12 = THDM.getParticle("Phi_12");
    Particle phi_1 = scalarboson_s("phi_1", THDM);
    Particle eta_1 = scalarboson_s("eta_1", THDM);
    Particle rho_1 = scalarboson_s("rho_1", THDM);
    eta_1->setSelfConjugate(true); // real field
    rho_1->setSelfConjugate(true); // real field
\end{lstlisting}
\end{framed}
Then making \marty\ apply the expansion in the Lagrangian:
\begin{framed}
\begin{lstlisting}
    Expr v_1 = constant_s("v_1");
    THDM.replace(Phi_11, phi_1());
    THDM.replace(Phi_12, (v_1 + rho_1() + CSL_I*eta_1()) / sqrt_s(2));
\end{lstlisting}
\end{framed}

\subsection{The $W^\pm$ particle}

For the sake of simplicity, the replacement procedure to apply
\begin{equation}
    W^\pm = \dfrac{W_1\mp iW_2}{\sqrt{2}}
\end{equation}
in \marty\ is not shown here. One can however find this exact procedure in the \href{https://marty.in2p3.fr/gettingStarted.html}{Getting Started} example available on the website.

\section{Rotating to the mass eigenstate basis}

\label{sec:rotation}
Gauge eigenstates after symmetry breaking are not eigenstates of the Hamiltonian and will thus not be the particles we may observe at colliders. To have the spectrum of the theory, fields must be rotated to the base in which their mass matrices are diagonal. In the 2HDM non diagonal mixings arise for pairs of fields as shown in Table \ref{tab:massbasis}, yielding 2-by-2 mass matrices.
\begin{table}[!h]
    \centering
    $\begin{array}{c|c|c}
        \text{Gauge eigenstates} & \text{Mass eigenstates} & \text{Angle} \\\hline
         (B^\mu, W_3^\mu)        & (A^\mu, Z^\mu) & \theta_W\\
         (\phi_1^\pm,\phi_2^\pm) & (G^\pm, H^\pm) & \beta \\
         (\eta_1,\eta_2)         & (G^0, A^0)     & \beta \\
         (\rho_1, \rho_2)        & (h^0, H^0)     & \alpha 
    \end{array}$
    \caption{Relations between gauge and mass eigenstates in the 2HDM, with the angle definition from one base to another. The massless Higgs degrees of freedom $G^\pm$ and $G^0$ are both found after a rotation of angle $\beta$.}
    \label{tab:massbasis}
\end{table}

Amongst the four mass matrices to diagonalize, three have zero determinant i.e. a zero eigenvalue: the photon $A^\mu$, $G^\pm$ and $G^0$. This property allows \marty\ to diagonalize them directly and in full generality without introducing overly complicated mixings and masses. For the massive states $(h^0, H^0)$ it is preferable to introduce by hand a mixing matrix\footnote{This procedure can be used for all matrices if the automated diagonalization, for zero determinant matrices, is less relevant e.g. introducing more complicated mixings.}, that will considerably lighten the resulting Lagrangian. Following the convention of Ref.~\cite{2HDM} the mixing matrix reads
\begin{equation}
    \left(\begin{array}{c}
         \rho _1  \\
         \rho_2 
    \end{array}\right) = 
    \left(\begin{array}{c c }
         -\sin\alpha & \cos\alpha  \\
         \cos\alpha  & \sin\alpha 
    \end{array}\right)\cdot
    \left(\begin{array}{c}
         h^0  \\
         H^0 
    \end{array}\right).
\end{equation}
This is done by \marty\ with the following commands:
\begin{framed}
\begin{lstlisting}
    Expr alpha = constant_s("alpha"); // Rotation angle
    THDM.rotateFields(
            {"rho_1", "rho_2"},  // Rotate from (rho_1, rho_2)
            {"h^0", "H^0"},          // to (h^0, H^0)
            {{-sin_s(alpha), cos_s(alpha)}, // Mixing matrix 
             {cos_s(alpha), sin_s(alpha)}},
            true); // Define this mixing as diagonalizing the masses
\end{lstlisting}
\end{framed}
When performing such a rotation \marty\ creates the new fields, applies the rotation, and diagonalizes mass terms by introducing symbolic masses that are considered to be independent parameters. Similarly one can rotate the other Higgs degrees of freedom\footnote{Applying the appropriate identities for the Higgs potential parameters \marty\ can diagonalize those degrees of freedom automatically as demonstrated later for the photon case. For simplicity those identities are not used here and the rotation must then be done manually.} as presented below.
\begin{framed}
\begin{lstlisting}
    Expr beta = constant_s("beta"); // Rotation angle
    THDM.rotateFields(
            {"eta_1", "eta_2"},  // Rotate from (eta_1, eta_2)
            {"G^0", "A^0"},          // to (G^0, A^0)
            {{cos_s(beta), sin_s(beta)}, // Mixing matrix 
             {sin_s(beta), -cos_s(beta)}},
            true); // Define this mixing as diagonalizing the masses
    // Same rotation for phi _1/2 -> G/H^+
\end{lstlisting}
\end{framed}
One may then automatically diagonalize the remaining zero-determinant matrices and rename particles\footnote{One cannot be certain in general about the order of the mass eigenstates in the new basis and must try to diagonalize the fields at least once to know it, e.g. which one of $B$ and $W^3$ gives the photon $A$.}:
\begin{framed}
\begin{lstlisting}
    // Finds and diagonalizes automatically the A_Y / W_3 matrix
    THDM.diagonalizeMassMatrices(); 
    THDM.renameParticle("A_Y", "A");
    THDM.renameParticle("W_3", "Z^0");
\end{lstlisting}
\end{framed}
For gauge fixing purposes it is better to link the Goldstone bosons $G^\pm$ and $G^0$ to their corresponding vector bosons $W^\pm$ and $Z^0$ through:
\begin{framed}
\begin{lstlisting}
    THDM.promoteToGoldstone("G^+", "W^+");
    THDM.promoteToGoldstone("G^0", "Z^0");
\end{lstlisting}
\end{framed}
Doing so will allow \marty\ to correctly fix the $W^\pm$ and $Z^0$ gauges (in particular changing the Goldstone bosons masses) when changed by the user. More details on those procedures together with more general ways to get the spectrum of a BSM model can be found in the \href{https://marty.in2p3.fr/doc/marty-manual.pdf}{MARTY manual}.

\section{Model completion}
\label{sec:completion}
\subsection{Lagrangian post-processing}

Once the main structural modifications have been applied, it is often necessary to perform some replacements of expressions to simplify the Lagrangian or to recover the standard conventions. In the 2HDM considered here we can introduce the electromagnetic coupling $e$, and replace the combination of VEVs and Yukawa by fermion masses (here presented only for the electron). This is shown below.
\begin{framed}
\begin{lstlisting}
    Expr g_Y = THDM.getScalarCoupling("g_Y"); // Getting the coupling from the model
    Expr g_L = THDM.getScalarCoupling("g_L");
    Expr theta_W = constant_s("t_W"); // Weinberg angle
    Expr e = constant_s("e"); // Electromagnetic constant
    Expr m_e = constant_s("m_e"); // Electron mass
    THDM.replace(g_Y, e / cos_s(theta_W));
    THDM.replace(g_L, e / sin_s(theta_W));
    THDM.replace(Y_e, sqrt_s(2) * m_e / v_1);
\end{lstlisting}
\end{framed}
Other clever replacements would further simplify this model but the few presented here are sufficient to demonstrate the general principles that can be applied in all cases.

\subsection{Model refreshing}
Once the model is completely defined, one must call the \lstinline!refresh()! function before using the model for the calculations, e.g.
\begin{framed}
\begin{lstlisting}
    THDM.refresh();
\end{lstlisting}
\end{framed}
This will allow \marty\ to clean the Lagrangian, merge identical terms, and more importantly to read the resulting mass terms of the Lagrangian to set once and for all the masses of the particles in diagrams\footnote{Masses and widths can be changed at any time by users, but this procedure is automated and should be called only once, after model building and before further calculations.}. This function also detects Weyl fermions connected by a mass term of the form
\begin{equation}
    \mathcal{L}_{mass}\ni -m_\psi\left(\bar{\psi}_L\psi_R + \bar{\psi}_R\psi_L\right),
\end{equation}
and simplifies the Lagrangian accordingly, replacing same (resp. opposite) terms for left- and right-handed parts by scalar (resp. pseudo-scalar) couplings using projection relations
\begin{align}
    &P_R+P_L=1,\\
    &P_R-P_L=\gamma^5.
\end{align}
\marty\ will in the 2HDM case perform the following simplifications automatically:
\begin{align}
    &\mathcal{L}\ni \frac{2}{3}ie(\bar{u}_L\gamma^\mu u_L + \bar{u}_R\gamma^\mu u_R)A_\mu\mapsto \frac{2}{3}ie \left(\bar{u}\gamma^\mu u\right) A_\mu,\\
    &\mathcal{L}\ni \frac{im_e}{v_1}\sin\beta(\bar{e}_Le_R - \bar{e}_Re_L)A^0\mapsto \frac{im_e}{v_1}\sin\beta \left(\bar{e}\gamma^5e\right)A^0,
\end{align}
which is in agreement with Eq.~(16) of Ref.~\cite{2HDM} for the lepton specific 2HDM. Other identities also hold for the other fermions and $h^0$, $H^0$ and $G^0$ couplings. Those simplifications are of a great importance as they can considerably reduce the number of Feynman diagrams for a given process, which grows exponentially with the number of interaction vertices.

During the whole process, the model can be displayed in the terminal using the instruction presented below.
\begin{framed}
\begin{lstlisting}
    cout << THDM << endl;
\end{lstlisting}
\end{framed}
Abbreviations introduced during the call of the \lstinline!refresh()! function are accessed with 
\begin{framed}
\begin{lstlisting}
    DisplayAbbreviations();    
\end{lstlisting}
\end{framed}
In the 2HDM case presented here one may see in the abbreviations the $W^\pm$ and $Z^0$ masses introduced by \marty\ during the call of \lstinline!diagonalizeMassMatrices()!
\begin{align}
    M_W &= \frac{ev}{2\sin\theta_W},\\
    M_Z &= \frac{ev}{2\sin\theta_W}\sqrt{1 + \tan^2\theta_W} = \dfrac{M_W}{\cos\theta_W},
\end{align}
with $v\equiv \sqrt{v_1^2+v_2^2}$ and $e=\sin\theta_W g_L=\cos\theta_W g_Y$. The masses derived by \marty\ correspond indeed to the well-known electroweak theory identities given respectively in Eqs. (29.10) and (29.11) in Ref.~\cite{Schwartz}.

After a model is built in \marty, it can be used to calculate amplitudes, cross-sections and Wilson coefficients~\cite{marty,Uhlrich:2020aaj}.

\section{Conclusion and perspectives}

We demonstrated how to build a model from scratch with \marty\ using the example of a 2HDM-like model. We preferred to keep this write-up as brief, simple and pedagogical as possible, therefore some relevant parts such as the flavour structure have been skipped. There is no limitation in \marty\ preventing us to generate more complex models and one can find in particular the full 2HDM (types I to IV) together with other (B)SM models in \marty's source code as explained in the {\it'Built-in models'} chapter of the manual. 

\begin{figure}[h!]
    \centering
    \includegraphics[width=0.67\linewidth]{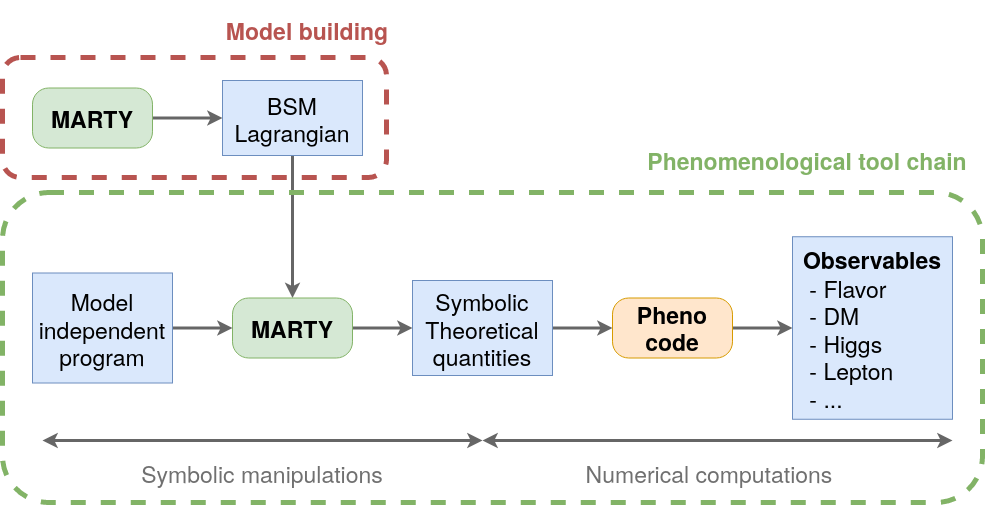}
    \caption{Theoretical tool chain using \marty\ as symbolic computer program doing calculations automatically, to feed phenomenological tools to obtain values for observables. The final output of \marty\ is symbolic theoretical quantities (amplitudes, squared amplitudes and Wilson coefficients at tree level or at one loop), in the form of C++ codes evaluating them numerically given a set of parameters.}
    \label{fig:toolchain}
\end{figure}

As mentioned in the Introduction, model building is the only task left once a phenomenological tool chain has been established, as presented in Figure \ref{fig:toolchain}. Such a phenomenological tool chain, in green, is mostly model independent and can be implemented only once. It requires the implementation of the initial \marty\ program that calculates specific quantities such as squared amplitudes or Wilson coefficients, and an interface with a phenomenological code to obtain observables from those quantities. Work is in progress in that direction using SuperIso \cite{superiso1, superiso2, superiso3} for the automated calculation of flavour observables in general BSM theories using Wilson coefficients calculated by \marty\ at the one-loop level, and SuperIso Relic \cite{relic1, relic2, relic3} for dark matter observables using squared tree-level amplitudes.

One can see that \marty\ has two distinct objectives: Calculate theoretical quantities at the one-loop level, but also help physicists to build BSM models. An important part of the development work has been and will still be dedicated to this task, making model building with \marty\ as simple and general as possible. Many topics on this aspect were not addressed here such as flavour groups, flavour symmetry breaking, gauge invariance, hermiticity, tensor manipulations in a Lagrangian, automated spectrum generation, etc. Detailed explanations and examples can be found in the \href{https://marty.in2p3.fr/doc/marty-manual.pdf}{manual}.

\bibliographystyle{h-physrev5}  
\bibliography{references}  

\begin{thebibliography}{10}

\bibitem{marty}
G.~Uhlrich, F.~Mahmoudi, and A.~Arbey, arXiv:2011.02478.

\bibitem{2HDM}
G.~Branco {\it et~al.},
\newblock Physics Reports {\bf 516}, 1–102 (2012).

\bibitem{dynkin}
P.~Cvitanović,
\newblock Group theory,
\newblock Princeton University Press, 2008, ISBN-9780691202983.

\bibitem{Schwartz}
M.~D. Schwartz,
\newblock {\em {Quantum Field Theory and the Standard Model}} (Cambridge
  University Press, 2014).

\bibitem{Uhlrich:2020aaj}
G.~Uhlrich, F.~Mahmoudi, and A.~Arbey,
\newblock arXiv:2011.06558.

\bibitem{superiso1}
F.~Mahmoudi,
\newblock Comput. Phys. Commun. {\bf 178}, 745 (2008), arXiv:0710.2067.

\bibitem{superiso2}
F.~Mahmoudi,
\newblock Comput. Phys. Commun. {\bf 180}, 1579 (2009), arXiv:0808.3144.

\bibitem{superiso3}
F.~Mahmoudi,
\newblock Comput. Phys. Commun. {\bf 180}, 1718 (2009).

\bibitem{relic1}
A.~Arbey and F.~Mahmoudi,
\newblock Comput. Phys. Commun. {\bf 181}, 1277 (2010), arXiv:0906.0369.

\bibitem{relic2}
A.~Arbey and F.~Mahmoudi,
\newblock Comput. Phys. Commun. {\bf 182}, 1582 (2011).

\bibitem{relic3}
A.~Arbey, F.~Mahmoudi, and G.~Robbins,
\newblock Comput. Phys. Commun. {\bf 239}, 238 (2019), arXiv:1806.11489.

\end{thebibliography}

\end{document}